\documentclass[numberedappendix,letterpaper]{emulateapj}
\usepackage{amstext}
\usepackage{apjfonts}
\usepackage{amsmath}
\usepackage{amssymb}
\usepackage[colorlinks=true,linkcolor=blue,citecolor=blue]{hyperref}
\usepackage{natbib}

\begin{document}

\newcommand{\beq}{\begin{equation}}
\newcommand{\eeq}{\end{equation}}
\newcommand{\beqar}{\begin{eqnarray}}
\newcommand{\eeqar}{\end{eqnarray}}
\newcommand{\e}{\epsilon}

\title{A Gravitational Double Scattering Mechanism for Generating High Velocity Objects} 

\author{Johan Samsing$^{1}$} 
\altaffiltext{1}{Dark Cosmology Centre, Niels Bohr Institute, University of Copenhagen, 
Juliane Maries Vej 30, 2100 Copenhagen, Denmark}

\begin{abstract} 
We present a dynamical model describing how halo particles can receive a significant energy kick from the merger
between their own host halo and a target halo. This is highly relevant for
understanding the growth of cosmological halos, and could especially provide an explanation for some high velocity objects.
The model we present includes a \emph{double scattering mechanism}, where a halo particle is given a significant energy kick
by undergoing two subsequent gravitational deflections during the merger.
The first deflection is by the potential of the target halo, whereas the
second is by the potential of the particle's original host halo.
The resultant energy kick arises because the two halos move relative to each other during the two deflections. To our knowledge, this
mechanism has never been characterized in this context before.
We derive analytically a halo particle's total kick energy, which is composed of energy
from the double scattering mechanism and energy release from tidal fields, as a function
of its position in its original host halo. In the case of a $1:10$ merger between two Hernquist halos,
we estimate that the presented mechanisms can generate particles with a velocity $\sim 2$ times the
virial velocity of the target halo measured at its virial sphere. This motivates us to suggest that the high velocity of the recently
discovered globular cluster HVCG-1 \citep{2014ApJ...787L..11C} can be explained by a head-on halo merger.
Finally, we illustrate the orbital evolution of particles outside the virial sphere of the target halo, by solving the equation of motion in
an expanding universe. We find a 'sweet spot' around a scale factor of $0.3-0.5$ for ejecting particles into large orbits, which easily
can reach beyond $\sim 5$ virial radii.
\end{abstract}


\maketitle

\section{INTRODUCTION}
Several high velocity objects on seemingly unbound orbits have
been observed, ranging from stellar objects \citep{2014ApJ...787...89B, 2014ApJ...785L..23Z},
supernovae (SNe) \citep{2003AJ....125.1087G, 2011ApJ...729..142S} and
gamma ray burst (GRBs) \citep{2013ApJ...769...56F, 2014MNRAS.437.1495T, 2014arXiv1401.7851B} to more extended systems like
globular clusters (GCs) \citep{2011ApJ...730...23P, 2014ApJ...787L..11C} and dwarf galaxies \citep{2007ApJ...670L...9M, 2007ApJ...662L..79C}.
In many of these cases the origin of the velocity kick is unknown, but
several mechanisms have been suggested. One is binary-single interactions where the
binding energy of a binary is dynamically released into a third object, which thereby can escape with high
velocity \citep{1975MNRAS.173..729H}.
These interactions are believed to have a non-negligible chance of happening especially between
stellar objects \citep{1993ApJ...415..631S, 2009MNRAS.396..570G}
and stars encountering either single or binary black
hole (BH) systems \citep{1988Natur.331..687H, 2003ApJ...599.1129Y, 2006ApJ...653.1194B}.
Several observations indicate in fact that stellar interactions with the supermassive black hole (SMBH) at the
center of our galaxy, is a likely explanation for some local high
velocity stars \citep{2005MNRAS.363..223G, 2012ApJ...754L...2B}.
More extended objects like GCs are probably not kicked by BH binary interactions, due to the high
probability for disruption, however the outcome from such an interaction is still uncertain \citep{2014ApJ...787L..11C}.
Dark matter (DM) subhalo interactions on the other hand,
is able to kick extended objects up to $\sim 2$ times the virial velocity of the host halo without major disruptions,
as indicated by numerical simulations \citep{2007MNRAS.379.1475S, 2009ApJ...692..931L}.
High velocity stars can also arise from isolated binaries if the heavier member undergoes a violent mass loss, a
channel first suggested by \cite{1961BAN....15..265B} to explain the high number of "run away" O-B stars.
More exotic kick mechanisms for describing hostless stellar remnants, pulsars, and possible hyper velocity BHs have been suggested as well,
from the role of asymmetric GW radiation \citep{1973ApJ...183..657B, 1983MNRAS.203.1049F, 1989ComAp..14..165R, 1995CeMDA..62..377P, 2002ApJ...579L..63D} to the asphericity of supernovae explosions \citep{1996AIPC..366...25B, 2007ApJ...655..416B, 2012ARNPS..62..407J}.

Unbound particles have also been discussed from a cosmological perspective. Recent studies \citep{2013JCAP...06..019B} 
illustrate that $\sim 10\%$ of all the DM at the virial radius is in fact unbound.
Luminous matter with no specific host halo has also been observed in especially galaxy clusters, a component known
as intra cluster light (ICL). This has been
extensively studied both through observations
\citep[e.g.][]{1951PASP...63...61Z, 2012A&A...537A..64G, 2014A&A...565A.126P} and numerically \citep{2004MNRAS.355..159W},
and is believed to be a direct consequence of the dynamical evolution of galaxies including tidal stripping
and mergers \citep{1996Natur.379..613M}.
Theoretical attempts have also been made to understand the final distribution of particles in DM halos. This includes
models from spherical collapse \citep[e.g.][]{1985ApJS...58...39B, 2010arXiv1010.2539D} to statistical
mechanics \citep[e.g.][]{1957SvA.....1..748O, 1967MNRAS.136..101L, 1992ApJ...397L..75S, 2005NewA...10..379H, 2010ApJ...722..851H, 2013MNRAS.430..121P}.
Especially concerning the unbound and high velocity component, recent work
by \cite{2009ApJ...707L..22T, 2009MNRAS.397..775J, 2014arXiv1405.6725C}
show that high velocity particles in mergers are likely generated through rapid mean field changes in the potential. This was also noted by
\cite{2009ApJ...691L..63A} who further proposed a direct connection to the observed population of high velocity B-type stars.

Data from upcoming
surveys like \emph{LSST}\footnote{http://www.lsst.org/lsst/} and especially \emph{Gaia}\footnote{http://sci.esa.int/gaia/}
will in the near future also measure positions and velocities for more than $\sim 150$ million stars with unprecedented precision.
This not only offers unique possibilities for mapping out the current Milky Way potential and
its past evolution \citep[e.g.][]{1999A&A...348L..49Z, 2012ApJ...760....2P, 2014arXiv1405.6721P, 2014IAUS..298..207S}, but will also
make it possible to make detailed studies of the past dynamical interactions \citep{2005MNRAS.363..223G}. A central question could here be
if the Milky Way in its past had a SMBH binary dynamically interacting
with the environment. Detections of high velocity objects are here again playing a central role.

In this paper we present a new dynamical mechanism for explaining how halo particles
gain a significant energy kick during the merger between their initial host halo and a target halo.
It is well known that halo mergers produce an unbound
component \citep[e.g.][]{1992ApJ...400..460H, 2009ApJ...707L..22T, 2014arXiv1405.6725C}, but no clear dynamical
explanation has been given yet. In this work we seek to give such an explanation.
Besides a well understood energy change from tidal
fields (also present in stellar disruption events \citep{1994ApJ...422..508K}), we show that
an additional mechanism play a significant role in changing the energy of the halo particles.
The new mechanism we present is a double scattering mechanism,
where a given particle receives a significant energy kick by undergoing two deflections during the merger.
We derive analytically the energy kick for two merging Hernquist halos \citep{1990ApJ...356..359H}, but the idea of the
mechanism is not limited to this scenario. For instance, we note that a very similar mechanism has been
described within heavy nuclei interactions where an electron can be ejected into the
continuum (unbound orbit) or captured by a passing nucleus (dynamical capture) by
undergoing a double collision\footnote{The quantum-mechanical solution to this interaction was not found until 1955 \citep{1955PhDT.........8D},
due to the fascinating fact that the second Born term is here
dominating over the first because of the double scattering nature, or two-step
process, of the problem.}\citep{1927RSPSA.114..561T, 1979RvMP...51..369S}.

The paper is organized in the following way:
In Section \ref{sec:overview_par_energy} we first give an introduction to the dynamical processes
playing a role in halo mergers which include tidal fields and our proposed
double scattering mechanism.
Section \ref{sec:General_Setup} describes our numerical simulations,
initial conditions, and the halo merger examples we consider in this paper.
The energy release from tidal fields is described in Section \ref{sec:Energy_From_Tidal_Fields}
and the double scattering mechanism is presented in Section \ref{sec:Kinematics_of_the_Double_Scattering}.
In both of these sections we derive the energy change of a given particle as a function of its position in its initial host halo.
In Section \ref{sec:Observational_Signatures} we shortly describe observable consequences
and show how a kick energy translates to an observable velocity excess.
In Section \ref{sec:cosmic_history_EPs} we explore how far a dynamically kicked particle can travel after leaving the
virial sphere of its target halo, by solving the equation of motion in an expanding universe.
Conclusions are given in Section \ref{sec:conclusion}.

\section{Energy of Particles During Halo Mergers}\label{sec:overview_par_energy}
The energy of individual halo particles can change significantly during a merger between their initial host halo ($H_{2}$) and a
larger target halo ($H_{1}$).
Some particles will lose energy and become bound to the target halo $H_{1}$, whereas some will gain energy and
escape with relatively high velocity. In this section we introduce the dynamical mechanisms responsible for
changing the energy of each individual particle initially bound to the incoming halo $H_{2}$.

\subsection{Dynamical Mechanisms}

\begin{figure}
\includegraphics[width=\columnwidth]{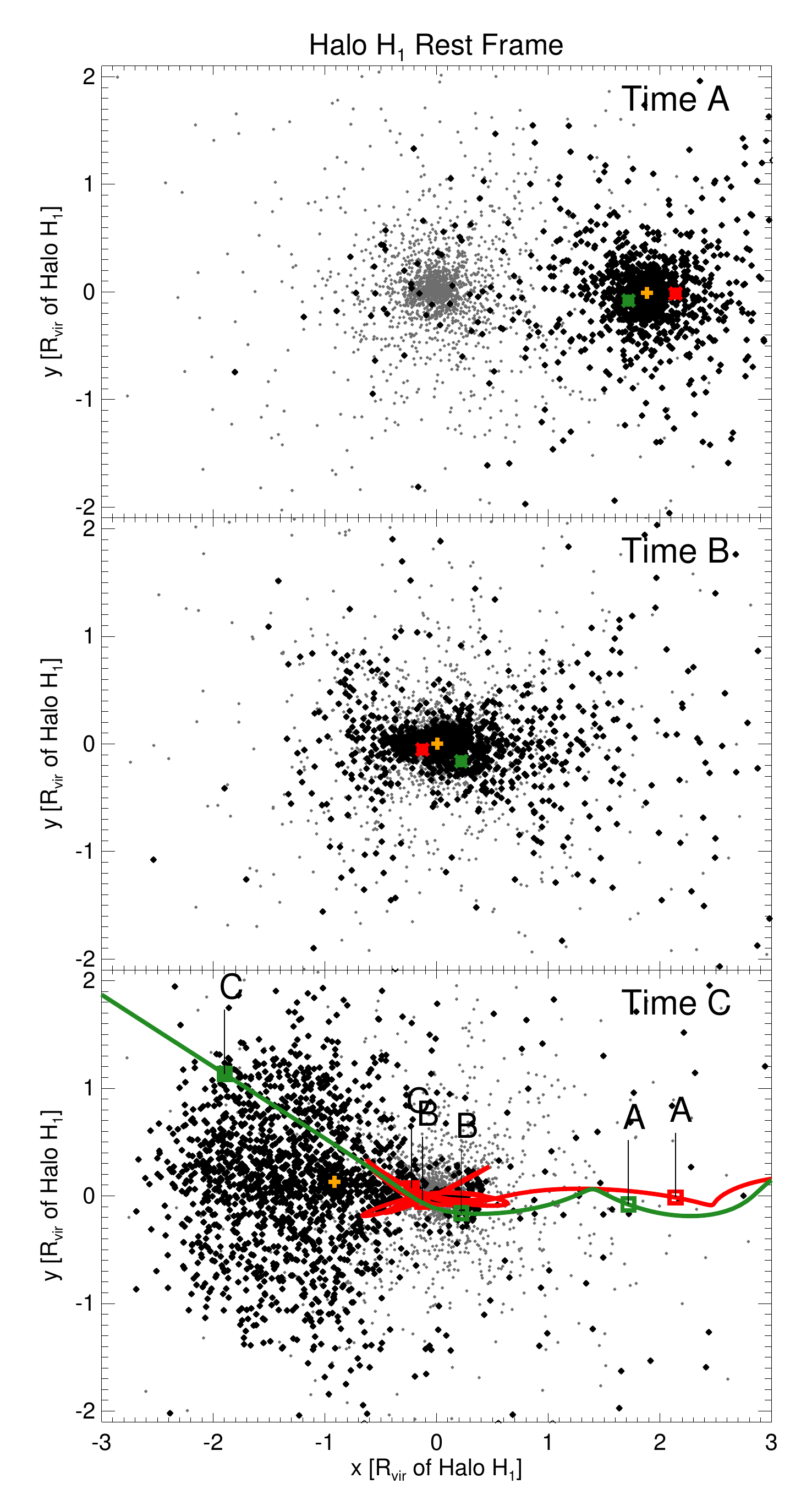}
\caption{Illustration of a $1:10$ merger between two DM halos merging with the escape velocity of the target halo.
The particles in the smaller incoming halo are shown in \emph{black}, while the particles of the larger target halo are
shown in \emph{grey}. The panels from \emph{top} to \emph{bottom} show three different times (A,B,C) of the merger. In the bottom panel
the full trajectories of two selected particles are also shown. The \emph{green} particle gains a positive energy kick during the merger and
is thereby escaping the system, whereas the \emph{red} particle loses energy and becomes bound to the target halo.
The \emph{orange} symbol shows a particle which is located within $5\%$ of the virial radius of the smaller incoming
halo at all times prior to the merger.
This illustrates a luminous galactic component. The merger clearly separates
the three highlighted particles in both position and velocity.
As described in Section \ref{sec:overview_par_energy}, this separation can be explained by two separate
dynamical processes: The first involve tidal fields (Section \ref{sec:Energy_From_Tidal_Fields})
and the second is our proposed double scattering mechanism (Section \ref{sec:Kinematics_of_the_Double_Scattering}).
Properties of the green particle are shown in Figure \ref{fig:change_in_v_E} and \ref{fig:particle_orbits_pos_RFH2} .}
\label{fig:particle_orbits_pos_RFH1}
\end{figure}

We first consider Figure \ref{fig:particle_orbits_pos_RFH1},
which shows an N-body simulation of a merger between two DM halos.
The incoming halo $H_{2}$ approaches from the right on a radial orbit with a velocity equal to the escape velocity of the target halo.
The orange symbol shows a particle which at all times
prior to the merger is located within $5\%$ of the incoming halo's virial radius.
This symbol can therefore represent a luminous galactic component \citep{2013ApJ...764L..31K}.
On the figure is also highlighted the orbits of two particles: The green particle receives a
positive energy change through the merger and can thereby escape, whereas the red particle gets bound
as a result of a negative energy change.
The differences in final energy between the orange, green, and red particles arise
due to a series of dynamical mechanisms, which to first order can be separately considered. Each of these
change the energy of the particles as described in the following.

The first energy change arises because the potential of the target halo $H_{1}$ is not constant across the profile of the incoming halo $H_{2}$.
As a result, the particles in $H_{2}$ located on the side closest to $H_{1}$ have less energy than the particles located on the far side.
This energy difference increases as the distance between the two halos decreases, i.e., a subject particle will gradually gain or lose energy
as the two halos approach each other. This continues until $H_{2}$ is tidally disrupted by the tidal field of $H_{1}$.
We denote the final energy change from this process by $\Delta{E}_{TF}$, where $TF$ is short for tidal field.

The second energy change arises from our proposed double scattering mechanism,
where a given particle is first deflected by $H_{1}$ and then subsequently by $H_{2}$.
This generates a change in energy because the two halos move relative to each other during
the two deflections. We denote the energy change from this process by $\Delta{E}_{DS}$, where $DS$ is short for double scattering.
To our knowledge, this contribution has not been characterized before and will therefore be the main topic of this paper.

The third and last energy change happens after the double scattering, as the particles are moving away from $H_{2}$ along their new
orbits. As the particles are climbing out of the potential of $H_{2}$ their velocity decreases, which results in an energy change in the frame of $H_{1}$
due to the relative motion between $H_{1}$ and $H_{2}$.
For particles escaping the merger remnant the energy change will be negative, as we will illustrate.
We denote the final energy change from this process by $\Delta{E}_{esc}$.

\subsection{Total Energy Change}

The dynamical mechanisms we consider in this work can, to first order, be considered separately and
not affecting the motion of a given particle at the same time. The
total energy change $\Delta{E}_{tot}$ a given particle will experience, can therefore be expressed
by the sum of the individual energy contributions
\begin{equation}
\Delta{E}_{tot} \approx \Delta{E}_{TF} + \Delta{E}_{DS} + \Delta{E}_{esc}.
\end{equation}
For particles receiving a high energy kick the first two terms will usually be positive and the last negligible.
In this paper we therefore focus on calculating the contribution from $\Delta{E}_{TF}$ and $\Delta{E}_{DS}$.
A numerical example of how the three energy terms individually change the total energy of a particle during a merger
is shown in the bottom panel of Figure \ref{fig:change_in_v_E}. The details of this figure will be described later.

\begin{figure}
\includegraphics[width=\columnwidth]{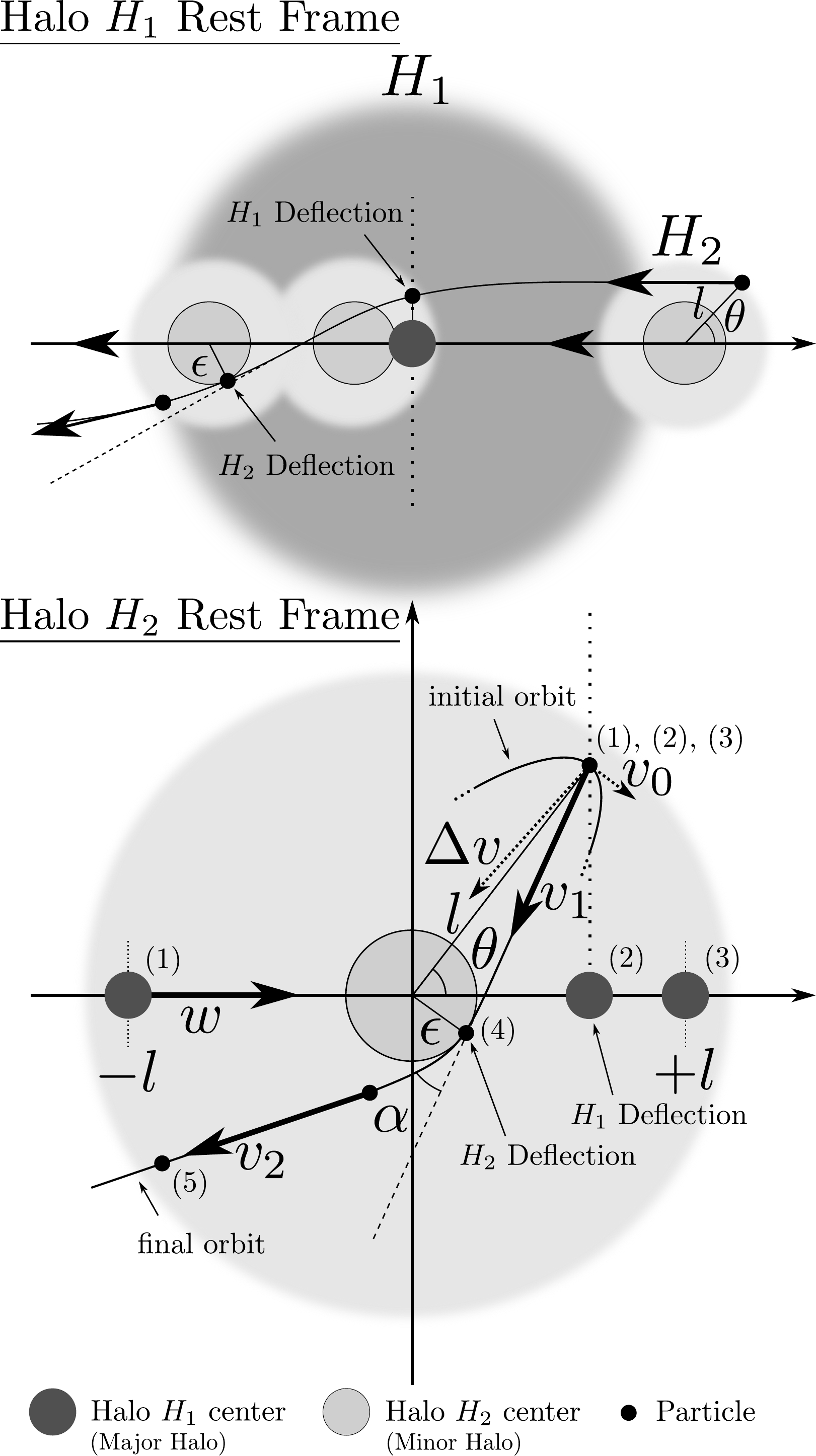}
\caption{
Schematic illustration of a particle (\emph{black dot}) gaining energy during the merger between its
original host halo $H_{2}$ (\emph{light grey}) and a
target halo $H_{1}$ (\emph{dark grey}).
The \emph{top plot} shows the orbital trajectory
of the particle in the rest frame (RF) of $H_{1}$, while the \emph{bottom plot} shows the trajectory in the RF of $H_{2}$.
As illustrated, the particle undergoes two separate deflections during its orbit:
The first is by the momentarily
dominating potential of $H_{1}$, whereas the second is by the potential of its original host halo $H_{2}$.
The deflection by $H_{1}$ results in a velocity kick $\mathbf{\Delta{v}}$ of the particle in the RF of $H_{2}$, as illustrated
in the bottom plot. The energy of the particle is conserved during this deflection in the RF of $H_{1}$, but not in the RF of $H_{2}$.
Due to the induced $\mathbf{\Delta{v}}$,
the particle now travels through $H_{2}$, where it scatters off the central parts of $H_{2}$ at
a peri-center distance $\sim \epsilon$. The energy of the particle during this deflection is conserved in
the RF of $H_{2}$, but not in the RF of $H_{1}$, because $H_{1}$ and $H_{2}$ are moving relative to each other. As a result
of the two deflections, the particle gains an energy kick $\Delta{E}_{DS}$ in the RF of $H_{1}$.
The numbers from (1-5) on the two paths represent simultaneous positions of the particle and $H_{1}$.
}
\label{fig:scattering_illustration}
\end{figure}

\section{General Setup}\label{sec:General_Setup}
We model the two merging halos $H_{1}$ and $H_{2}$ by
Hernquist (HQ) profiles \citep{1990ApJ...356..359H}, with an anisotropy parameter $\beta=0$. In this case the mass profile is given by
\begin{equation}
M_{i}(r) = M_{i}\frac{(r/a_{i})^{2}}{(1+r/a_{i})^{2}},
\end{equation}
and the corresponding gravitational potential by
\begin{equation}
\Phi_{i}(r) = -\frac{GM_{i}}{a_{i}}\frac{1}{(1+r/a_{i})},
\label{eq:HQ_Epot}
\end{equation}
where $M_{i}$ is the total mass of halo $i$, $r$ is the distance from the halo center, $M_{i}(r)$ is the mass enclosed by $r$,
$\Phi(r)$ is the potential at distance $r$ and $a$ is a characteristic scale radius.
In the following, we occasionally use units of $a_{1}$ and we use a prime to denote this, e.g., $x' \equiv x/a_{1}$.
We also find it useful, to write down the radial velocity between $H_{1}$ and a particle
moving in its potential on a radial orbit
\begin{equation}
w^{2}(r) = -2\Phi_{1}(r) + w^{2}(0) + 2\Phi_{1}(0),
\label{eq:w_relative}
\end{equation}
where $w(r)$ is the radial velocity of the particle at distance $r$. We will
use this relation to calculate the relative velocity between halo $H_{1}$ and the incoming halo $H_{2}$.
The estimate for $w(r)$ in the above equation \eqref{eq:w_relative} ignores the effect from dynamical friction,
which causes $H_{2}$ to lose orbital energy by exchanging momentum with the surrounding
particles in $H_{1}$ \citep{1943ApJ....97..255C}.
Dynamical friction actually plays a minor
role in our case because $\Delta{E}_{DS} \propto w$, as we describe
in Section \ref{sec:Kinematics_of_the_Double_Scattering}, but for now
we ignore it to simplify the analysis.
For the following analyzes, we further assume that the two halos merge with zero impact parameter and that the mass of the target
halo $H_{1}$ is much larger than the incoming halo $H_{2}$, i.e. $M_{1} \gg M_{2}$. This mass hierarchy
is relevant for the growth of cosmological halos, that are believed to build hierarchically by
hundreds of minor mergers \citep{2010MNRAS.406.2267F}.

The N-body simulations presented throughout the paper are performed
using Gadget II \citep{2005MNRAS.364.1105S},
with the two HQ halos set up in equilibrium by Eddington's Method \citep{1916MNRAS..76..572E} using a well
tested code previously used for studying the anisotropy in halo
mergers \citep{2012JCAP...07..042S, 2012JCAP...10..049S}. The halo concentration $c_{i} \equiv R_{i,\text{vir}}/a_{i}$ is set to $5$ for both
halos and the virial radius $R_{i,\text{vir}}$ is calculated by requiring the density inside the halo to be $200$ times the mean
density of the universe at redshift $z=2$ \citep{2002ApJ...568...52W, 2010gfe..book.....M}. 
We fix the merger mass ratio at $1:10$ for all simulations and the incoming halo $H_{2}$ is set to have zero energy relative
to the target halo $H_{1}$, which corresponds to a velocity at infinity $w_{\infty} = 0$.
These initial conditions are typical in a cosmological perspective \citep{2012MNRAS.423.3018P}, however
a wide range of both encounter velocities and impact parameters are seen in full cosmological simulations \citep{2011MNRAS.412...49W}.

\begin{figure}
\includegraphics[width=\columnwidth]{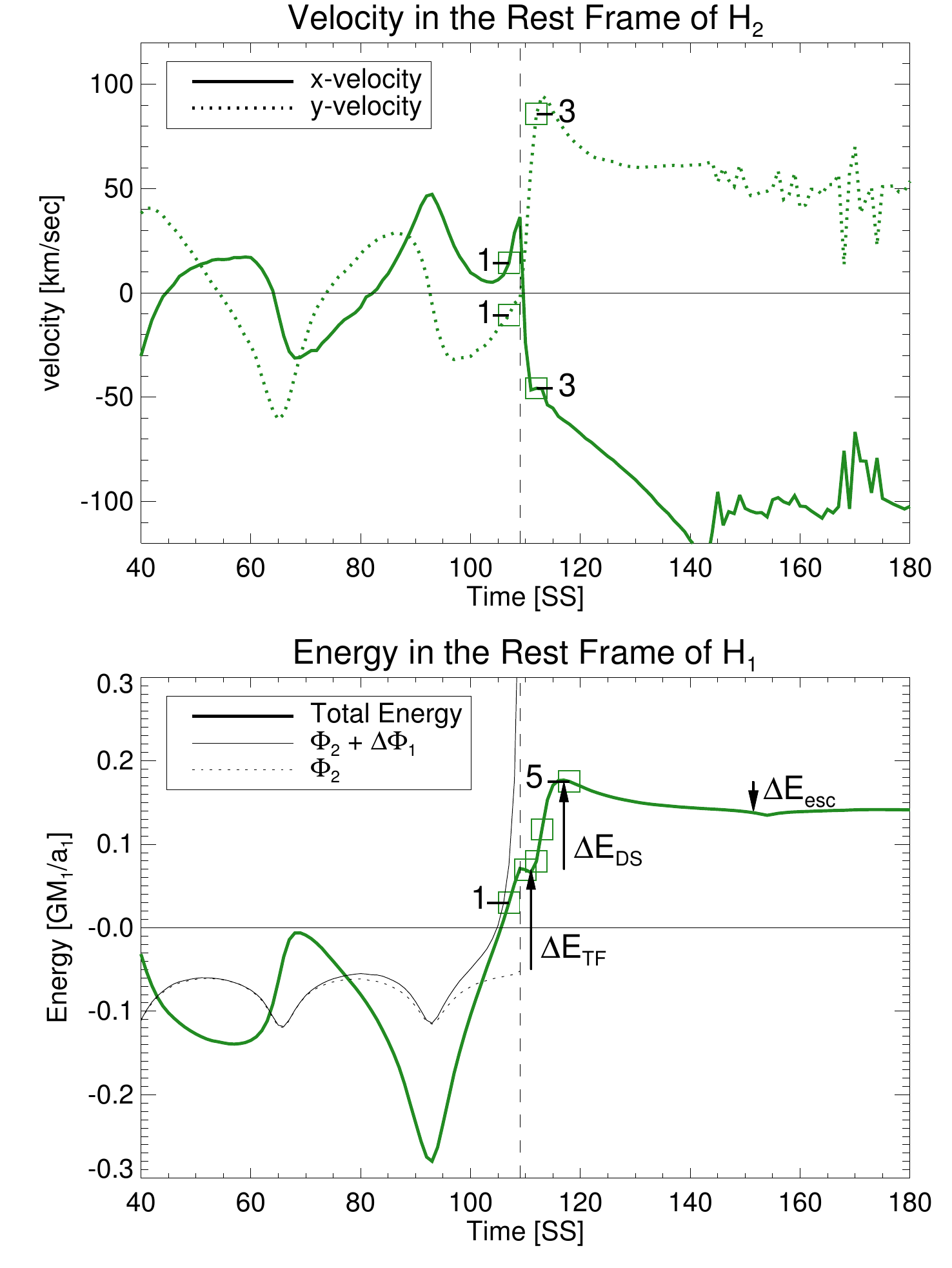}
\caption{
Velocity and energy as functions of time for a particle gaining a positive energy through the merger.
The particle is the same as shown in Figures \ref{fig:particle_orbits_pos_RFH1} and \ref{fig:particle_orbits_pos_RFH2}.
The vertical \emph{dashed line} indicates the time when the two merging halos pass each other.
\emph{Top:} Horizontal and vertical velocity of the particle in the RF of its original host halo $H_{2}$ as a function of time.
The numbered \emph{squares} (1,3) indicate when $H_{2}$ passes $H_{1}$ from $-l$ to $+l$
as illustrated in Figure \ref{fig:scattering_illustration}. As seen, the particle gains a significant
velocity kick $\mathbf{\Delta{v}}$ during this passage, as discussed in Section \ref{sec:velkick_from_H1}. 
\emph{Bottom:} Energy of the particle in the RF of $H_{1}$ as a function of time.
The first energy increase ($\Delta{E}_{TF}$) is due to the variation of $\Phi_{1}$ across $H_{2}$, as
discussed in Section \ref{sec:Energy_From_Tidal_Fields}, whereas the second energy increase ($\Delta{E}_{DS}$)
is generated through our proposed double scattering mechanism, discussed in Section \ref{sec:Kinematics_of_the_Double_Scattering}.
The gradual energy decrease ($\Delta{E}_{esc}$) at later times is happening, because the particle
is dragged back as it travels out of $H_{2}$, which itself moves relative to $H_{1}$.
Comparing with Figure \ref{fig:particle_orbits_pos_RFH2}, we see that the second increase happens
when the particle undergoes its 'second deflection' by $H_{2}$. This is in complete agreement with our double scattering model.
}
\label{fig:change_in_v_E}
\end{figure}

\section{Energy From Tidal Fields}\label{sec:Energy_From_Tidal_Fields}

The first energy change the particles in $H_{2}$ experience is from tidal fields.
This change arises, because the particles in $H_{2}$ all have the same bulk velocity $w(r)$, but
experience different values of $\Phi_{1}$ due to their different spatial positions in $H_{2}$.
The difference in $\Phi_{1}$ across $H_{2}$ increases as the two halos approach each other and
the particles are therefore released with a wide spread in energy at the time $H_{2}$ tidally disrupts. This scenario is very similar
to stellar disruption events (see e.g. \cite{1994ApJ...422..508K}).
In this section we derive an estimate for the energy change $\Delta{E}_{TF}$ a given particle in $H_{2}$
will experience from this process, as a function
of the particle's position in $H_{2}$ just prior to the merger.

\subsection{Evolution of Particle Energy Before Merger}
We consider a particle located in $H_{2}$, with orbital velocity $\mathbf{v}$ and polar position $l,\theta$ measured in the
rest frame (RF) of $H_{2}$. The configuration is illustrated in Figure \ref{fig:scattering_illustration}.
The energy of the particle in the RF of $H_{1}$ before the merger, is given by
\begin{equation}
E(l,\theta,r) = \frac{1}{2}w(r)^{2} + \Phi_{1}(r) + \frac{1}{2}v^{2} + \Phi_{2}(l) + \mathbf{w}{\cdot}\mathbf{v} + \Delta{\Phi_{1}}(l,\theta,r),
\label{eq:E_par_RF_H1}
\end{equation}
where $r$ is the distance between $H_{2}$ and $H_{1}$, $w$ is the corresponding relative velocity
and $\Delta{\Phi_{1}}$ is the difference between the value of $\Phi_{1}$ at the position of the
center of mass (CM) of $H_{2}$ and the particle, respectively.
The first two terms equal the CM energy of $H_{2}$ in the RF of $H_{1}$, where the next two terms
equal the energy of the particle in the RF of $H_{2}$,
i.e., the sum of the first four terms remains approximately constant, as the two halos approach each other.
The fifth term $\mathbf{w}{\cdot}\mathbf{v}$ is by contrast oscillating between positive and negative values as the particle orbits $H_{2}$.
Energy can be released from this term, but the contribution is random and does not simply add to the other energy contributions
for reasons we will not discuss here.
The only term that changes the energy of the particle in a constructive way is the last term $\Delta{\Phi_{1}}$, which is given by
\begin{equation}
\begin{split}
\Delta{\Phi_{1}}	(l,\theta,r)		& =  \frac{\Phi_{1}(0)}{1+ \sqrt{r'^2 + l'^2 + 2l'r'\text{cos}(\theta)}} - \frac{\Phi_{1}(0)}{1+r'}, \\
  						& \approx \Phi_{1}(0)\frac{l'\text{cos}(\theta)}{(1+r')^{2}},\ \ r' \gg l'.
\end{split}
\label{eq:delta_phi_TF}
\end{equation}
This illustrates that the change in energy of the particle due to the variation of $\Phi_1$ across $H_{2}$,
scales to linear order as $\sim l\text{cos}(\theta)/r^2$, i.e., particles in $H_{2}$ located on the side closest to $H_{1}$
lose energy as $H_{2}$ approaches $H_{1}$, whereas particles on the other side instead gain energy.

The energy contribution from the $\Delta{\Phi_{1}}$ term is seen in the bottom panel of Figure \ref{fig:change_in_v_E}, which shows
the energy of the green particle from Figure \ref{fig:particle_orbits_pos_RFH1} as a function of time. 
One can see that the $\Delta{\Phi_{1}}$ term does not contribute when $r$ is large (at early times), but as $r$ decreases and
becomes comparable to $l$,
the $\Delta{\Phi_{1}}$ term clearly increases and the total energy of the particle therefore increases as well.
This energy increase continues, until the particle tidally detaches from $H_{2}$ and starts to move completely under the
influence of $H_{1}$ (just before the vertical dashed line). The moment at which this happens,
can be estimated by comparing the tidal force exerted on the particle $F_{tid}$ by $H_{1}$ with
the binding force $F_{bin}$ by $H_{2}$ \citep{2006MNRAS.366..429R}.
These force terms are simply given by $F_{tid} = GM_{1}(d)/d^2 - GM_{1}(r)/r^2$ and $F_{bin} = GM_{2}/l^2$, where
$d$ denotes the distance from $H_{1}$ to the particle.
By defining the force ratio $\delta_{TF} \equiv F_{tid}/F_{bin}$, one can now relate $\delta_{TF}$ and the position of the particle in $H_{2}$
to a corresponding distance between the two halos $R_{TF}$. In the case of two HQ halos we find to linear order,
\begin{equation}
R'_{TF} \approx \delta_{TF}^{-1/3} \left[2l'\text{cos}(\theta)(a'_{2} + l')^{2}M_{1}/M_{2}\right]^{1/3} - 1, \ \ r' \gg l'.
\label{eq:disrupt_radius}
\end{equation}
If $\delta_{TF}=1$, then the corresponding $R_{TF}$ will be the standard definition of the tidal radius.

\subsection{Resultant Energy From Tidal Fields}
\begin{figure}
\includegraphics[width=\columnwidth]{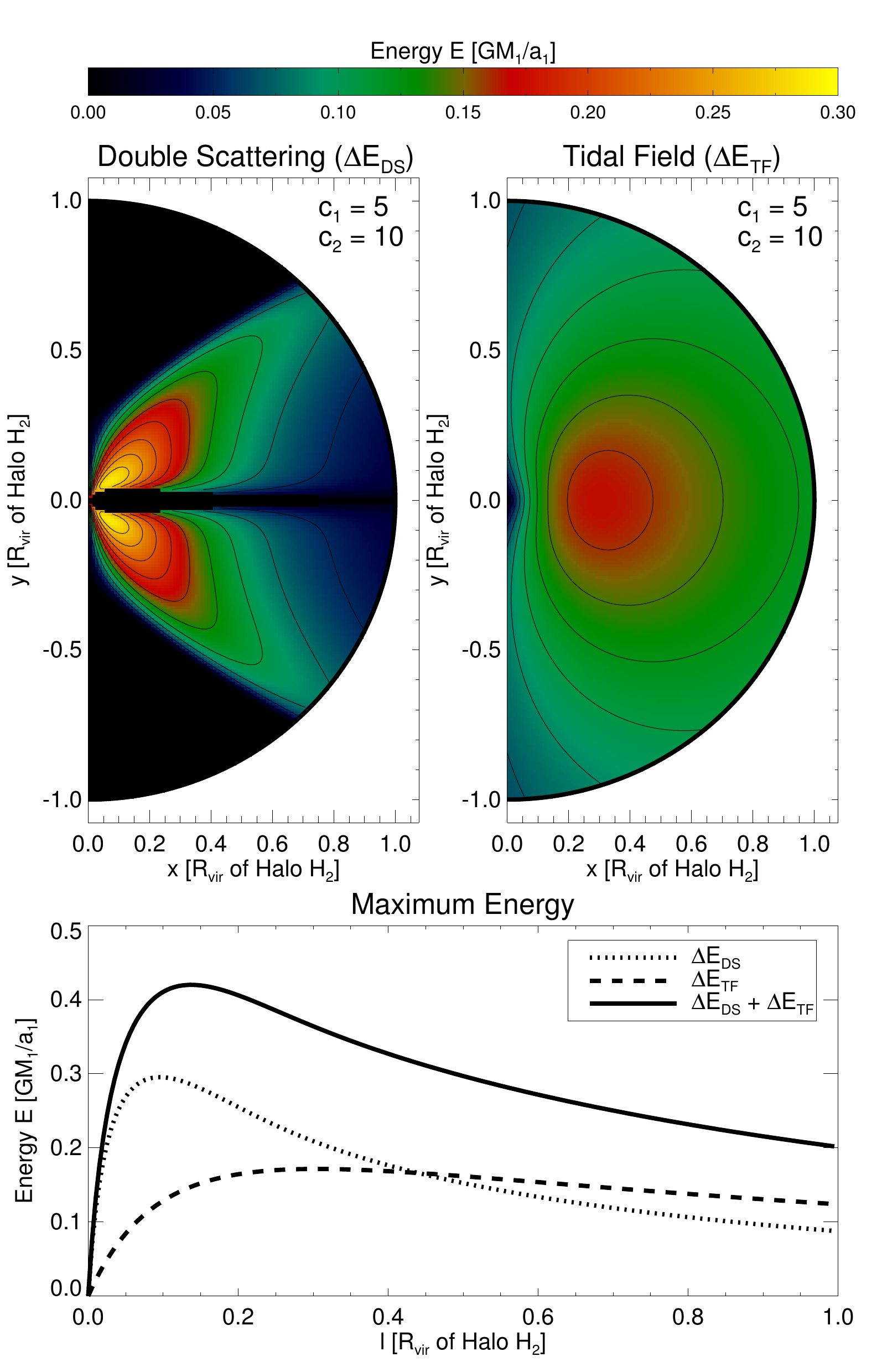}
\caption{
Dynamical kick energy generated by tidal fields ($\Delta{E}_{TF}$, section \ref{sec:Energy_From_Tidal_Fields})
and by our proposed double scattering
mechanism ($\Delta{E}_{DS}$, Section \ref{sec:Kinematics_of_the_Double_Scattering}),
as a function of particle position in $H_{2}$ just prior to the merger.
The results are for a $1:10$ head-on merger between two HQ halos with concentrations, $c_{1}=5$ and $c_{2}=10$,
passing each other with the escape velocity of $H_{1}$.
\emph{Top}: Contour plots showing our theoretical calculated kick energy
as a function of the position of the particle in $H_{2}$ just prior to the merger.
The \emph{left} plot shows the contribution from tidal fields $\Delta{E}_{TF}$,
where the right plot shows the contribution from the double scattering
mechanism $\Delta{E}_{DS}$ (using $1.5\Delta{v_{y}}$ to correct for the known
bias as explained in Section \ref{sec:Vertical_Kick_Velocity}).
The two plots only show the right hand side ($x_{p}>0$) of the incoming halo $H_{2}$,
which in this example is approaching $H_{1}$ from right to left.
Particles with $x_{p}>0$ will gain the illustrated energy, whereas particles with $x_{p}<0$ instead will lose this energy. This follows
trivially from our analytical estimates.
\emph{Bottom}: The maximum kick energy as a function of the distance $l$ of the particle
from the CM of $H_{2}$. As one can see, the maximum kick energy from the two mechanisms is around $\sim 0.4\Phi_{1}(0)$.
This will lead to particles with a velocity around two times the virial velocity of the target halo $H_{1}$, as discussed in Section
\ref{sec:Observational_Signatures}.
}
\label{fig:allowed_dE_space}
\end{figure}

The resultant energy change $\Delta{E_{TF}}$ of the particle induced by the tidal field of $H_{1}$, is given by evaluating
the potential energy difference $\Delta{\Phi_{1}}$ (equation \eqref{eq:delta_phi_TF}) at distance $R_{TF}$ (equation \eqref{eq:disrupt_radius})
where the particle tidally detaches from $H_{2}$,
\begin{equation}
\Delta{E_{TF}}(l,\theta,R_{TF}) \approx \Delta{\Phi_{1}}(l,\theta,r=R_{TF}(\delta_{TF})).
\end{equation}
A fair agreement with numerical simulations is found when $\delta_{TF} \approx 3-5$. However, we
also find slight deviations which primarily are caused by the difficulties in defining a representative $R_{TF}$.
To do a better estimation, one needs to include the
possibility for the particle to detach gradually, but this is highly non trivial.
A gradual detachment is, e.g., seen in the energy evolution of
the particle shown in Figure \ref{fig:change_in_v_E}.
For clarity, we therefore instead report the energy change $\Delta{E_{TF}}$ the particle has received, when the
two halos are separated by the distance $r=l$ (denoted by '1' in Figure \ref{fig:change_in_v_E}).
The corresponding energy is both accurately determined and
representative for the resultant energy change $\Delta{E_{TF}}$
induced by $\Phi_{1}$. We find this to be true for the majority of the particles in our simulation.
The right panel in Figure \ref{fig:allowed_dE_space} shows $\Delta{E_{TF}}(l,\theta,r=l)$ as a function of the
position of the particle in $H_{2}$. We see that the change in energy is estimated to be around $0.1-0.2\Phi_{1}(0)$
and the maximum kick is given to particles just behind the center of $H_{2}$.
The tidal field contribution $\Delta{E_{TF}}$ is also illustrated and discussed in Figure \ref{fig:change_in_v_E}.

\begin{figure}
\includegraphics[width=\columnwidth]{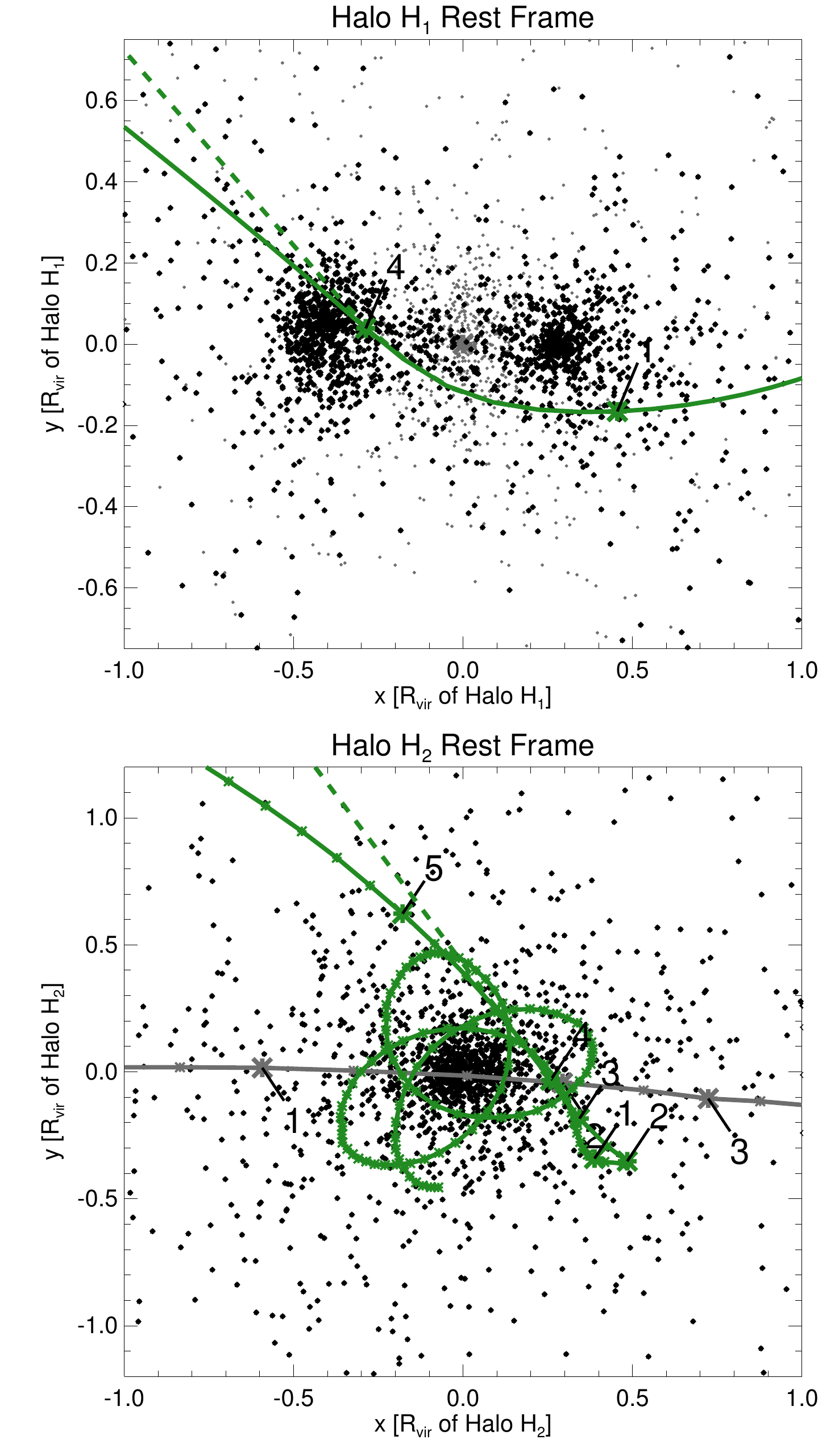}
\caption{
Orbital trajectory of a halo particle (\emph{green}) gaining a significant energy kick from the merger
between its own initial host halo $H_{2}$ (\emph{black})
and a target halo $H_{1}$ (\emph{grey}). The particle is the same as the green one shown in Figure \ref{fig:particle_orbits_pos_RFH1}.
The \emph{green solid} lines show the full orbit of the particle, whereas the \emph{green dashed} lines show the orbit if the particle
is not undergoing its 'second deflection' by $H_{2}$.
This second deflection directly leads to the energy kick $\Delta{E}_{DS}$
generated by the double scattering mechanism, as described in Section \ref{sec:analytical_model}.
The dashed line would therefore lead to no energy increase from this mechanism. 
The \emph{numbers} refer to five important moments as illustrated in Figure \ref{fig:scattering_illustration}.
\emph{Top}: Orbit of the particle in the RF of $H_{1}$. The halo particles of $H_{2}$ are plotted
at time $1$ (right halo) and $4$ (left halo), respectively.
\emph{Bottom}: Orbit of the particle in the RF of $H_{2}$. The horizontal \emph{grey line} shows the
orbit of the target halo $H_{1}$ that moves from left to right.
The smaller \emph{stars} on the orbits indicate equal time intervals. The angle between the solid and the dashed line is
denoted by $\alpha$, and is calculated in equation \eqref{eq:alpha_deflec}.
The corresponding time dependent velocity and energy of the particle are shown in Figure \ref{fig:change_in_v_E}.
}
\label{fig:particle_orbits_pos_RFH2}
\end{figure}

\section{Energy From the Double Scattering Mechanism}\label{sec:Kinematics_of_the_Double_Scattering}
The second energy change the particles in $H_{2}$ experience during the merger is generated by the double scattering mechanism.
In this section we describe the kinematics of the mechanism and derive an analytical solution for the resulting
kick energy $\Delta{E}_{DS}$. As seen in Figure \ref{fig:change_in_v_E}, the energy contributions from tidal fields and the double scattering
mechanism are of the same order. The mechanism is therefore playing an important role for
how energy is distributed in halo mergers.

\subsection{Origin of the Double Scattering Kick Energy}\label{sec:Origin_of_the_kick_Energy}
The double scattering mechanism is a process where a
particle is gravitationally deflected two times during
the merger between its own host halo $H_{2}$ and a target halo $H_{1}$.
The first deflection is by the potential of $H_{1}$, which momentarily dominates as the two halos overlap,
whereas the second is by the potential of $H_{2}$, which can dominate after the two halos have passed each other.
We refer to the deflection by $H_{1}$ as the
'first deflection' and the subsequent deflection by $H_{2}$ as the 'second deflection'.
The two merging halos are moving relative to each other during the merger, so the two deflections are therefore happening in
two different velocity frames.
The energy of the particle during each deflection is conserved in the frame of deflection, 
but because the two frames move relative to each other, a deflection in one frame can result in an energy change in the other.
In our case, the deflection by $H_{2}$ changes the velocity of the particle by an amount ${\delta}v$ along the
motion of the two merging halos. The particle energy is constant in the frame of $H_{2}$, but in the frame of $H_{1}$
the energy changes by an amount $\sim ({\delta}v + w)^2 - w^2 \sim {\delta}vw$. This is the contribution from the double scattering mechanism,
denoted by $\Delta{E}_{DS}$. A schematic illustration is shown in Figure \ref{fig:scattering_illustration}, where a numerical
example is shown in Figure \ref{fig:particle_orbits_pos_RFH2}. In the following, we calculate the details of this double scattering process.

\subsection{Analytical Model}\label{sec:analytical_model}
We consider a particle initially bound to $H_{2}$, with orbital velocity $\mathbf{v_{0}}$ and polar position
$l,\theta$ measured in the RF of $H_{2}$ just prior to the merger. For reaching an analytical solution for the double scattering kick energy
$\Delta{E}_{DS}$, we now work from the orbital
picture shown in Figure \ref{fig:scattering_illustration}, which serves to approximate the full orbital trajectory of the
particle. Following this picture, we first model the velocity kick $\mathbf{\Delta{v}}$ the particle receives relative to
the CM of $H_{2}$ from its 'first deflection' by $H_{1}$. We then use this kick velocity to model the orbit of the particle
through $H_{2}$, where it undergoes its 'second deflection' by the mass of $H_{2}$
enclosed by radius $\epsilon$. This deflection rotates the velocity vector of the particle by an angle $\alpha$,
resulting in a velocity change $\delta{v}$ along the motion of the merging halos. From this deflection, we then calculate
the resultant energy change $\Delta{E}_{DS} \sim \delta{v}w$, as described in Section \ref{sec:Origin_of_the_kick_Energy}.
The components of this model will be calculated in the sections below for two merging HQ halos.

\subsubsection{The 'First Deflection' by Halo $H_{1}$}\label{sec:velkick_from_H1}

The particle receives a velocity kick $\mathbf{\Delta{v}}$ relative to $H_{2}$, because
the CM of $H_{2}$ and the particle experience different accelerations
during the merger. A numerical example is shown in the top panel of Figure \ref{fig:change_in_v_E}.
We can analytically estimate $\mathbf{\Delta{v}}$ in the impulsive limit, where one assumes
the particle is not moving during the
encounter \citep[e.g.][]{1983ApJ...268..342H, 1985ApJ...295..374A, 1991CeMDA..52..263C, 1999ApJ...511..625F}.

Using a coordinate system where
$H_{1}$ is moving along the x-axis and the CM of $H_{2}$ is located at $x=0$, the kick can now be estimated by
\begin{equation}
\mathbf{\Delta{v}}	 \approx \int_{-T}^{+T} \mathbf{{a}}(t) dt = \int_{-R}^{+R} \frac{1}{w(x)} \frac{GM_{1}(d)}{d(x)^{3}} {\mathbf{d}(x)}dx,
\label{eq:general_eq_dv}
\end{equation}
where $\mathbf{a}$ is the acceleration the particle experiences due
to $H_{1}$, $\mathbf{d} = (d_{x},d_{y})$ is the separation vector between the particle
and the CM of $H_{1}$, $d = {\mid}\mathbf{d}{\mid}$ is its magnitude
and $w$ is the relative velocity between $H_{1}$ and $H_{2}$.
The distance $d$ is simply given by $d^{2}=(x-x_{p})^2 + y_{p}^2$, where
$x_{p} = l \text{cos}(\theta)$ and $y_{p} = l \text{sin}(\theta)$ are the x and y coordinates of the particle in the frame of $H_{2}$, respectively.
In this work, we model the orbit of the particle by assuming it is not moving during the passage of $H_{1}$ from
$-l,+l$, as illustrated in Figure \ref{fig:scattering_illustration}. The horizontal kick velocity $\Delta{v_{x}}$ is therefore
calculated by setting $R=l$. The vertical kick velocity $\Delta{v_{y}}$ is not sensitive to $R$ in the same way, and is therefore
calculated from just using $R=\infty$ to simplify the expressions.
The horizontal and the vertical components
of the kick velocity will be calculated below.

\subsubsection{Horizontal Kick Velocity $\Delta{v_{x}}$}

The horizontal kick velocity $\Delta{v_{x}}$ is found by integrating
equation \eqref{eq:general_eq_dv} from $-l$ to $+l$ using $d_{x} = x-x_{p}$,
\begin{equation}
\Delta{v}_{x} \approx \frac{\Phi_{1}(0)}{w(l)}\left(\frac{1}{1+l'\sqrt{2}\sqrt{1-\text{cos}(\theta)}} - \frac{1}{1+l'\sqrt{2}\sqrt{1+\text{cos}(\theta)}}\right),
\label{eq:velkick_dvx_FULL}
\end{equation}
where we have assumed that $w$ equals $w(l)$ during the passage.
By comparing with equation \eqref{eq:HQ_Epot}, we see that the expression, except for the $1/w$ term, is exactly equal
to the difference in potential energy
of the particle between the initial configuration, where $H_{1}$ is at $-l$, and the final configuration, where $H_{1}$ is at $+l$.
This is consistent from the perspective of energy conservation, where the
particle must receive a kinetic energy kick to 'compensate' for the potential
energy difference $\Delta{\Phi_{-l,+l}}$. To illustrate this, we note that in
the RF of $H_{1}$ the kinetic energy of the
particle after the merger is $E_{kin}(l)\approx w(l)\Delta{v_{x}}$ and from energy conservation the kick must therefore be
$\Delta{v_{x}} \approx \Delta{\Phi_{-l,+l}}/w(l)$ as we
also find in equation \eqref{eq:velkick_dvx_FULL}.
The horizontal velocity kick is therefore
not due to a real dynamical deflection, but it arises purely from an energy difference. This difference can be calculated exactly,
and as a result our estimate for $\Delta{v_{x}}$ is also relatively accurate.
In practice, it is useful to approximate equation \eqref{eq:velkick_dvx} by
$\Delta{v}_{x}(\theta){\approx}\Delta{v}_{x}(0)\text{cos}(\theta)$, where $\Delta{v_{x}}(\theta)$ denotes
the solution including the full $\theta$ dependence.
Using this approximation we find
\begin{equation}
\begin{aligned}  
\Delta{v}_{x}	& \approx \frac{\Phi_{1}(0)}{w(l')}\frac{2l'}{1+2l'} \text{cos}(\theta).
\label{eq:velkick_dvx}
\end{aligned}
\end{equation}
In the limit where $H_{1}$ and $H_{2}$ pass through each other with the escape velocity of $H_{1}$ this
reduces to the simple form: $\Delta{v}_{x} \approx -\sqrt{2{\mid}\Phi_{1}(0){\mid}}{l'} \text{cos}(\theta) \sqrt{1+l'}/{(1+2l')}$.

\subsubsection{Vertical Kick Velocity $\Delta{v_{y}}$}\label{sec:Vertical_Kick_Velocity}

The vertical kick velocity $\Delta{v}_{y}$ arises
because the particle briefly follows an orbit in the potential of $H_{1}$, which momentarily dominates
as the two merging halos pass each other.
The velocity kick $\Delta{v}_{y}$ can therefore be
estimated from writing down the orbital solution for a particle, with encounter velocity $\sim w$ and impact
parameter $\sim l\text{sin}(\theta)$, moving in the HQ potential of $H_{1}$.
However, there are no analytical solutions for the majority of DM density profiles, including
the HQ profile \citep{2008gady.book.....B}, and we must therefore use the impulsive approximation presented in equation \eqref{eq:general_eq_dv}.
Assuming  the particle is only deflected by the mass of $H_{1}$ enclosed by a sphere of radius $r={\mid}{l\text{sin}(\theta)}{\mid}$, and
using $d_{y} = y_{p}$, we find
\begin{equation}
\Delta{v}_{y}	 \approx \frac{\Phi_{1}(0)}{w(x'_{p})} \frac{2y_{p}'}{(1+{\mid}y_{p}'{\mid})^{2}},
\label{eq:velkick_dvy}
\end{equation}
where we have assumed that $w$ equals ${w(x'_{p})}$ during the passage ($w$ at time '2' shown in Figure \ref{fig:scattering_illustration}).
In the limit where $H_{1}$ and $H_{2}$ pass through each other with the escape velocity of $H_{1}$, the above expression reduces to
$\Delta{v}_{y} \approx -\sqrt{2{\mid}\Phi_{1}(0){\mid}}{y_{p}'}\sqrt{1+x'_{p}}/{(1+{\mid}y_{p}'{\mid})^{2}}$.
In contrast to the horizontal kick $\Delta{v_{x}}$, the vertical kick $\Delta{v_{y}}$
arises from a real dynamical deflection,
which makes it hard to estimate precisely. 
By comparing with simulations, we find that our above estimation for $\Delta{v_{y}}$ is about a
factor of $\sim 1.5$ too low. One reason for this is that we only include the mass of $H_{1}$ enclosed by the radius $\sim {l\text{sin}(\theta)}$.
However, including the full HQ profile in the integration leads to a divergent result, which clearly illustrates the limits
of the impulsive approximation.

\begin{figure}
\includegraphics[width=\columnwidth]{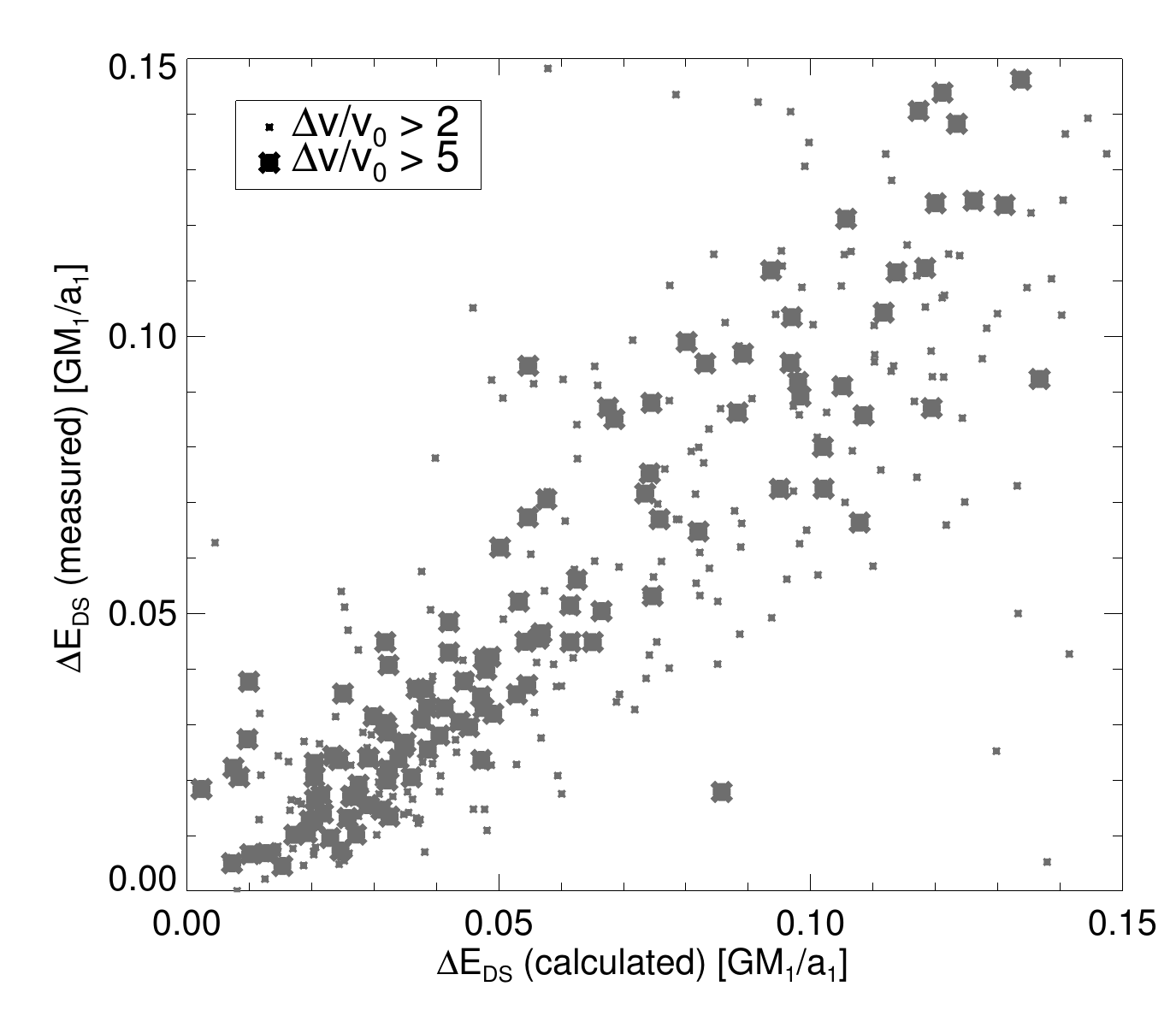}
\caption{
Comparison between our analytical calculation for the dynamical energy kick generated by the double scattering mechanism (x-axis) and
values measured from an N-body simulation (y-axis). The analytical estimate is done using equation \eqref{eq:ana_dE}, with measured values
for $\Delta{v}$ and $w(r_{\epsilon}')$ to completely focus on the mechanism itself.
The two symbol sizes indicate different thresholds between
the dynamical kick velocity, $\Delta{v}$, and the peculiar motion of the particle, $v_{0}$,
at the time of merger. As seen, our model successfully describes the kick energy from the double scattering mechanism.
}
\label{fig:Energy_parplot_scatter}
\end{figure}

\subsubsection{The 'Second Deflection' by Halo $H_{2}$}\label{sec:second_scattering_by_H2}

After receiving the velocity kick $\mathbf{\Delta{v}}$, the particle starts to move from its initial position $l,\theta$
towards the central region of $H_{2}$, where it undergoes a 'second deflection' by the mass of $H_{2}$ enclosed by radius $\epsilon$.
This changes the velocity vector of the particle from $\mathbf{v_{1}} = \mathbf{v_{0}}+\mathbf{\Delta{v}}$
to $\mathbf{v_{2}} = \mathbf{v_{1}}+\mathbf{\delta{v}}$.
To estimate the components of $\mathbf{v_{2}}$, we first calculate the impact parameter $\epsilon$ for the deflection by $H_{2}$, as
illustrated in Figure \ref{fig:scattering_illustration}.
Assuming $\left| \text{tan}(\theta) \Delta{v_{x}}/\Delta{v_{y}} \right|<1$ we find from simple geometry
\begin{equation}
\epsilon = \left| x_{p} \right| \frac{1-{\gamma}{\text{tan}(\left| \theta \right|)}}{\sqrt{1+\gamma^{2}}},
\end{equation}
where $\gamma \equiv \left| {\Delta{v_{x}}}/{\Delta{v_{y}}} \right|$.
Using the relation $\alpha \approx \delta{v}/\Delta{v}$ and equation \eqref{eq:general_eq_dv} to estimate $\delta{v}$,
we can now write down an expression for the deflection angle $\alpha$
\begin{equation}
\begin{aligned} 
\alpha 		& \approx  \frac{2GM_{2}(\epsilon)}{\epsilon \Delta{v}^2} = \frac{2{\mid}\Phi_{2}(0){\mid}}{\Delta{v}^2}\frac{(\epsilon/a_{2})}{(1+\epsilon/a_{2})^2},
\end{aligned}
\label{eq:alpha_deflec}
\end{equation}
assuming that the particle is only
affected by the mass of $H_{2}$ enclosed by $\epsilon$.
In the last equality we have inserted the HQ mass profile of $H_{2}$.
The deflection by $H_{2}$ conserves the length of the velocity vector of the particle in
the RF of $H_{2}$, but rotates $\mathbf{v_{1}}$ by the angle $\alpha$ into the new vector $\mathbf{v_{2}}$, which
therefore has coordinates given by
\begin{equation}
\begin{aligned} 
v_{2,x} 		& =  v_{1,x}\text{cos}\alpha + v_{1,y}\text{sin}\alpha \\
v_{2,y}		& =  v_{1,y}\text{cos}\alpha - v_{1,x}\text{sin}\alpha .
\end{aligned}
\end{equation}
The particle will only receive a positive energy kick if ${\mid}v_{2,x}{\mid} > {\mid}v_{1,x}{\mid}$, i.e.,
if the kick velocity $\mathbf{\Delta{v}}$ and deflection angle $\alpha$ fulfill the inequality
${\mid} \text{tan}(\alpha/2)\Delta{v_{x}}/\Delta{v_{y}}{\mid} < 1$ in
the limit where $\Delta{v} \gg v_{0}$.
From the definition of $\mathbf{\delta{v}} \equiv \mathbf{v_{2}} - \mathbf{v_{1}}$ we now find the change in
velocity due to the second deflection
\begin{equation}
\begin{aligned} 
{\mid}\delta{v_{x}}{\mid} \approx \left| \Delta{v_{y}}\alpha \right|  \\
{\mid}\delta{v_{y}}{\mid} \approx \left| \Delta{v_{x}}\alpha \right|
\label{eq:vel_diff_from_H2_scatter}
\end{aligned}
\end{equation}
where we have assumed that $\alpha \ll 1$ and that the kick velocity dominates the
motion of the particle along its new perturbed orbit, i.e. $v_{1}\approx \Delta{v}$.
The last assumption is necessary for the double scattering mechanism to work effectively.

\subsubsection{Resultant Energy From the Double Scattering Mechanism}\label{sec:Change_of_Energy}

To finally calculate the dynamical kick energy $\Delta{E}_{DS}$ of the particle,
we first assume that the second deflection by $H_{2}$
happens instantaneously, i.e., the velocity vector
of the particle changes from $\mathbf{v_{1}}$ to $\mathbf{v_{2}}$ at a single point.
This point occurs when the particle passes the center of $H_{2}$
at a distance $\sim \epsilon$, as shown in Figure \ref{fig:scattering_illustration}.
From this assumption it naturally follows that the potential energy of the particle
is approximately constant during the deflection, and the change in total energy will therefore
be dominated by the change in kinetic energy. The kick energy $\Delta{E}_{DS}$ can therefore be estimated by
\begin{equation}
\Delta{E}_{DS}(l, \theta) \approx \frac{1}{2}(\mathbf{v_{2}} + {\mathbf{w}(r'_{\epsilon})})^2 - \frac{1}{2}(\mathbf{v_{1}} + {\mathbf{w}(r'_{\epsilon})})^2 = {w(r'_{\epsilon})}\delta{v_{x}},
\label{eq:ana_dE}
\end{equation}
where $\delta{v_{x}}$ is the x component of the velocity change in the RF of $H_{2}$ given
by equation \eqref{eq:vel_diff_from_H2_scatter}, $r'_{\epsilon}$ is
the distance between $H_{1}$ and $H_{2}$ at the time the particle undergoes its second deflection by $H_{2}$
and ${w(r'_{\epsilon})}$ is the corresponding relative
velocity between $H_{1}$ and $H_{2}$.
In the limit where the two halos pass each other with the escape velocity of $H_{1}$, $r'_{\epsilon}$ is found by solving the differential equation
$w(r) = dr/dt = \sqrt{2{\mid}\Phi_{1}(r){\mid}}$. The solution for a HQ halo can be written in the
form $r'_{\epsilon} = ( 3{\Delta{t}}{\sqrt{2{\mid}\Phi_{1}(0){\mid}}}/{(2a_{1})} + (1+l')^{3/2})^{2/3}-1$,
where $\Delta{t} \approx l/\Delta{v}$ is the time from the first deflection by $H_{1}$ to the second deflection by $H_{2}$.
For a slightly more precise estimate one has to include dynamical friction, which impacts the estimation for $w$.
The friction will mainly play a role in slowing down the bulk of $H_{2}$ after the merger, i.e., the main change will be to the
value of ${w(r'_{\epsilon})}$. The correction will therefore factor out
in equation \eqref{eq:ana_dE}, which makes it easy to include in possible future studies.

The left panel in Figure \ref{fig:allowed_dE_space} shows our estimate for $\Delta{E}_{DS}$, given by equation \eqref{eq:ana_dE},
as a function of the position of the particle in $H_{2}$.
One can see that our model predicts that the particles which receive a positive
energy kick are all located in a cone with two wings pointing along the velocity of $H_{2}$.
Comparing with the energy kick generated by tidal fields $\Delta{E}_{TF}$ (illustrated in the right panel), we see that the double scattering mechanism
is actually likely to be the dominating kick mechanism for particles located near the center. This is in contrast to the outer parts,
where the tidal field contribution seems to be the dominating component. We confirmed this by numerical simulations.

A comparison between an N-body simulation and our analytical estimate for $\Delta{E}_{DS}$ is shown
in Figure \ref{fig:Energy_parplot_scatter}. The analytical calculation is done using equation \eqref{eq:ana_dE},
with numerical measured values for $\Delta{v}$ and $w(r_{\epsilon}')$, to completely isolate the prediction from
the double scattering mechanism itself.
The measured energy kick from the N-body simulation is here defined as the change in energy of the particle between the time
when $H_{2}$ leaves $H_{1}$ at distance $l$ (time '3') and the time when the particle leaves $H_{2}$ at distance $l$ (time '5').
As seen on the figure, we find good agreement despite the difficulties in both measuring and calculating the kick energy.

\section{Observational Signatures}\label{sec:Observational_Signatures}

The particles which have gained an energy kick through our presented mechanisms will have a relative high
velocity compared to the field, and therefore have the potential of being labeled as high velocity objects.
In this section, we illustrate how different the resultant velocities of the kicked
particles are, compared to the virialized particles bound to $H_{1}$. 

\subsection{Kick Velocity Relative to Virial Velocity}

The velocity of a particle moving on a radial orbit in the potential of the target halo $H_{1}$, is found from simple energy conservation
\begin{equation}
u_{r}(r) = \sqrt{2(E_{i} + \Delta{E} - \Phi_{1}(r))},
\label{eq:radvel_in_RF_H1_at_distR}
\end{equation}
where $u_{r}(r)$ is the radial velocity of the particle at distance $r$, $E_{i}$ is the initial
energy of the particle, $\Phi_{1}(r)$ is the radial dependent potential of $H_{1}$ and
$\Delta{E}$ is any additional energy contributions. We consider the case where $\Delta{E} = \Delta{E}_{DS} + \Delta{E}_{TF}$.
All quantities are defined in the RF of $H_{1}$.

\begin{figure}
\includegraphics[width=\columnwidth]{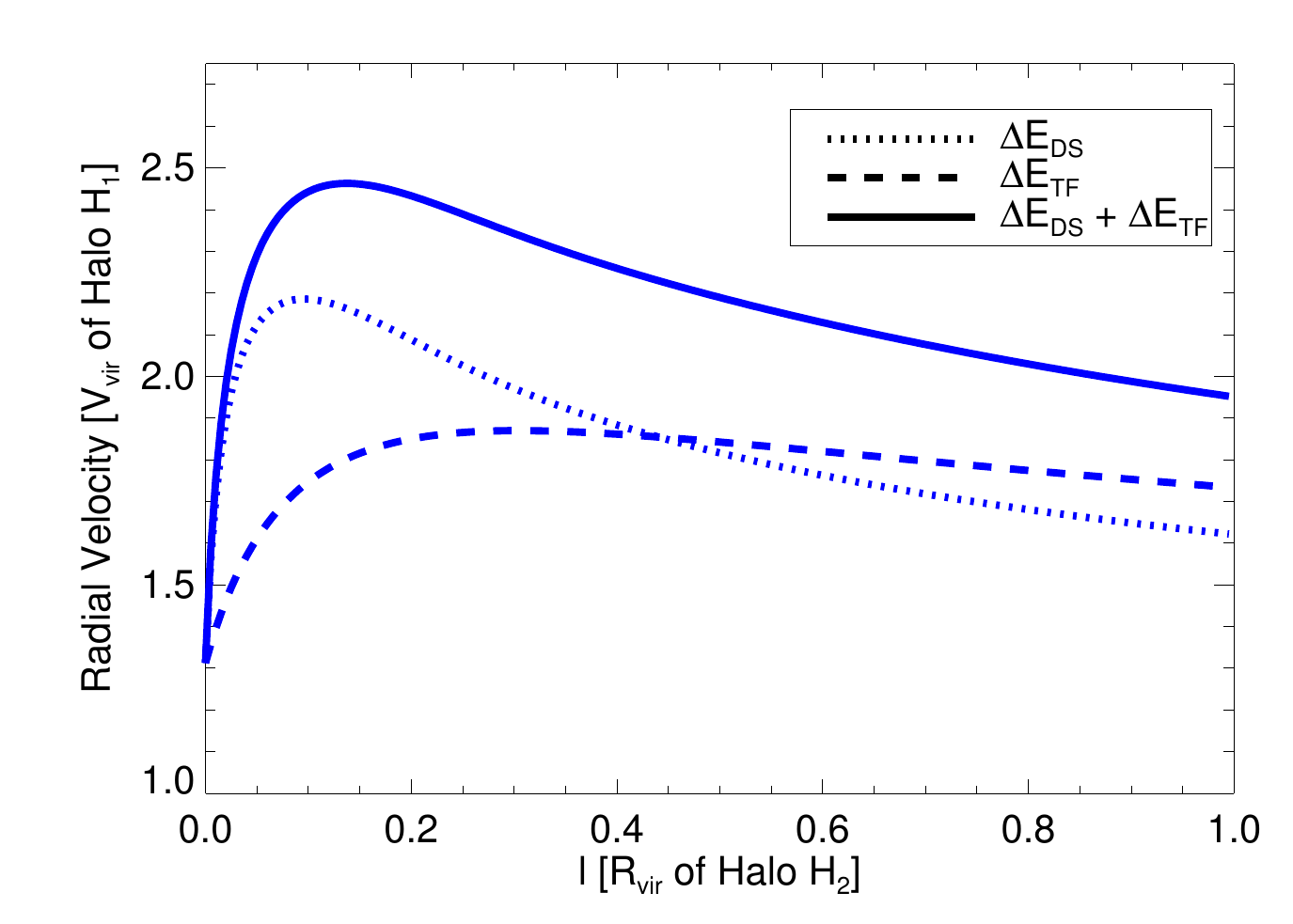}
\caption{
The maximum radial velocity evaluated at the virial sphere of $H_{1}$ (i.e. at distance $r=R_{1,\text{vir}}$)
for a particle kicked by our presented mechanisms, as a function of its radial position $l$ in $H_{2}$ just prior to the merger.
The merger configuration is the same as the one discussed in Figure \ref{fig:allowed_dE_space}.
The velocity curves are calculated from the maximum energy estimates shown in Figure \ref{fig:allowed_dE_space} using equation 
\eqref{eq:radvel_in_RF_H1_at_distR}. One can see that the dynamical mechanisms giving rise to $\Delta{E}_{TF}$
and $\Delta{E}_{DS}$ can create particles traveling with velocities $\sim 2V_{1,\text{vir}}$ at the virial sphere of their target halo.
}
\label{fig:max_dVRvir_at_Rinfty}
\end{figure}

Figure \ref{fig:max_dVRvir_at_Rinfty} illustrates our analytical estimate for the maximum radial velocity a dynamically kicked particle
can have at the virial sphere of $H_{1}$. The velocity is plotted in units of the virial velocity of $H_{1}$,
defined by $V_{1,\text{vir}} \equiv \sqrt{GM_{1,\text{vir}}/R_{1,\text{vir}}}$.
As seen on the figure, the energy release from tidal fields and the double scattering mechanism can lead to particles with a velocity
about $\sim 2V_{1,\text{vir}}$ at the virial radius of $H_{1}$. These particles will therefore clearly
stand out from the virialized part. For comparison, particles receiving no energy
kicks ($\Delta{E} = 0$) will, in our example, instead have a velocity $\sim 1.3V_{1,\text{vir}}$.
We also see that the maximum kick velocity is given to particles located
around $\sim 0.1-0.2 R_{2,\text{vir}}$ from the
center of $H_{2}$. Stars are typically located within $1-5 \%$ of the virial radius of
their host halo \citep{2013ApJ...764L..31K}, we therefore expect only the outer parts of a possible central galaxy
to be effectively kicked by our presented mechanisms.
The outer parts are usually populated by loosely bound stars and
stellar systems, such as GCs and dwarf galaxies \citep{2013MNRAS.428..389P}.
The GCs are the only of these objects which can be seen out to cosmological distances, due to their high number and
density of stars ($\sim10^{4}\ \text{pc}^{-3}$), which make them a potential observable tracer of
our presented mechanisms. We give an example of this in the section below.

\subsection{Is HVGC-1 Kicked Through a Halo Merger?}

The first detection of a high velocity globular cluster (HVGC-1) was recently reported by \cite{2014ApJ...787L..11C}.
This high velocity object was identified as a GC from spectroscopy, and \emph{uiK} photometry and was found between GC candidates collected
over several years by Keck/DEIMOS, LRIS and MMT/Hectospec \citep{2011ApJS..197...33S, 2012ApJ...748...29R}.
The GC is located in the Virgo Cluster
at a projected distance of $\sim 84$ kpc from M87, with a radial velocity relative to Virgo and M87 of
about $2100$ and $2300$ $\text{km} \ \text{s}^{-1}$,
respectively. The interesting question is now, how did this GC get this high velocity?
As discussed in the paper by the authors, the GC could have been kicked by a binary SMBH system located in the center of M87. However, it
is very uncertain whether a GC can survive this, due to the possibility of disruption. Subhalo interactions near M87 could also
be an explanation, but no subhalos have been observed in its close vicinity yet. The nature of the kick is therefore still unsolved. 

The GC could have been kicked by our presented dynamical mechanisms, i.e., first by tidal fields
and then by the double scattering mechanism, if it was initially bound to a DM halo merging nearly head-on with Virgo.
To receive the maximum kick energy, the GC must have been
located in the outskirts of its host galaxy just prior to the merger, which is not
an unlikely scenario \citep[e.g.][]{2014MNRAS.442.2165H, 2013MNRAS.428..389P}.
For a $1:10$ mass ratio, we have shown that this merger configuration can generate objects with a radial velocity
of $\sim 2$ times the virial velocity of the target halo at its virial radius.
In the case of Virgo, this would mean a velocity of
about $\sim 2 \times 1100 = 2200 \ \text{km} \ \text{s}^{-1}$ (The virial velocity of
Virgo is somewhere between $\sim 900-1300 \ \text{km} \ \text{s}^{-1}$ \citep{2011ApJS..197...33S}), which
is consistent with the observed value for HVGC-1.
We further note that HVGC-1 is observed to be hostless.
This also follows from our model, since the generated kick velocity of the GC quickly separates it from its initial host
galaxy. This is also seen in Figure \ref{fig:particle_orbits_pos_RFH1}.
Full numerical simulations can of course be used for exploring this in more detail, including the role
of encounter velocity, halo concentrations and mass profiles, impact parameter, and mass ratio.
We leave that for a future study.

\section{Orbits Outside the Virial Radius}\label{sec:cosmic_history_EPs}

An interesting final question, is now what the future orbits are of the dynamically kicked particles if they leave the virial sphere
of their target halo $H_{1}$ (also denoted the 'ejector halo' in the sections below).
We study this, by solving the equation of motion for particles
moving under the influence of the gravitational force of their ejector halo and the expanding cosmological background.

\begin{figure}
\includegraphics[width=\columnwidth]{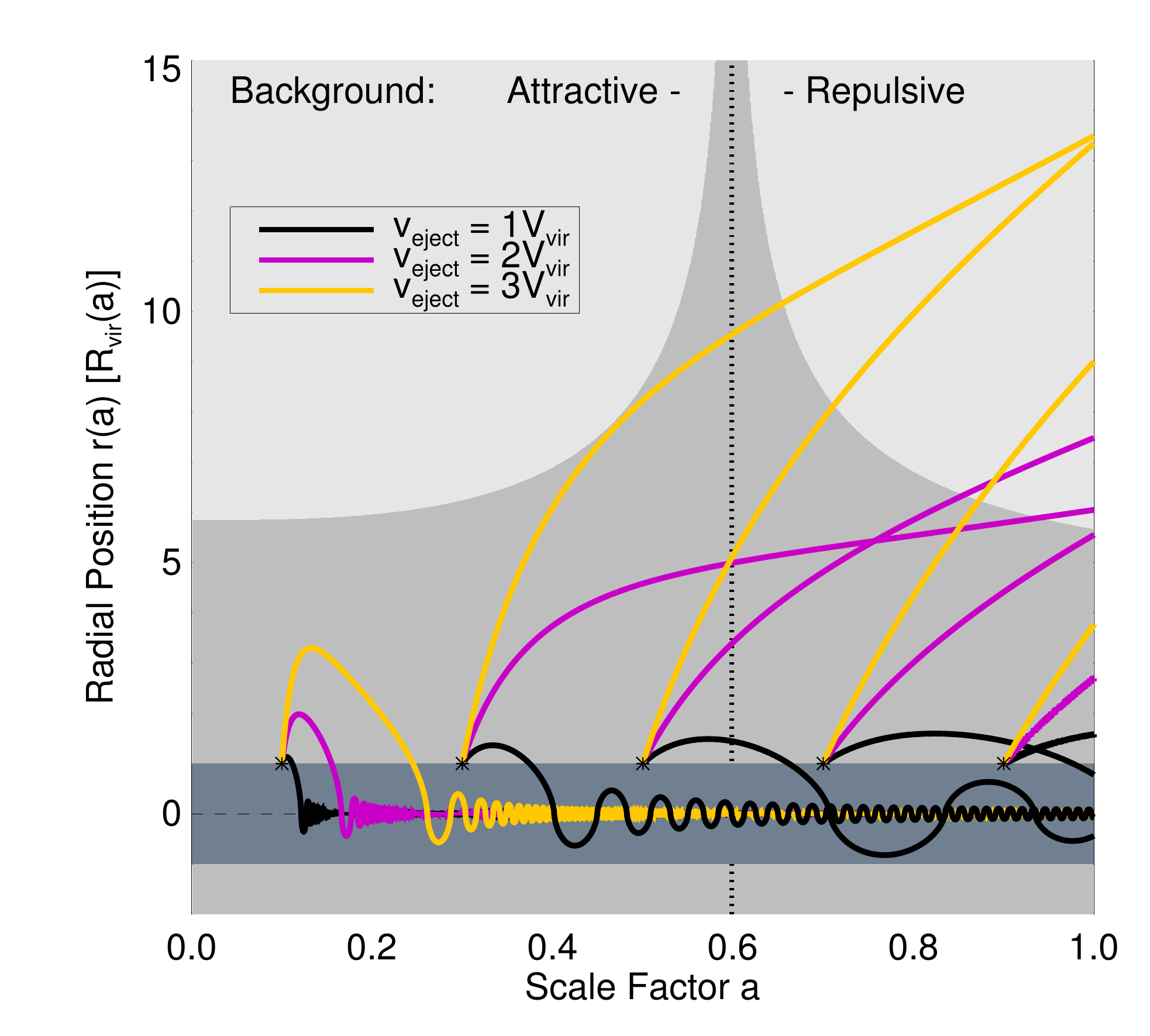}
\caption{Radial position (\emph{solid lines}) as a function of scale factor $a$ for particles
leaving the virial sphere of $H_{1}$ with velocity $v_{eject}$ at different times (\emph{black stars}).
The \emph{dark grey} area centered around zero shows the region inside the virial
sphere of $H_{1}$.
The \emph{grey} area shows the region where the
gravitational force from the halo dominates over the force from the expanding background.
In the \emph{light grey} region the force from the background dominates.
The vertical \emph{dotted line} indicates the
time when the force from the expanding background changes from being attractive (left side) to
repulsive (right side), i.e., when $\Omega_{m,0}a^{-3}=2\Omega_{\Lambda,0}$.
One can see that there exists a 'sweet spot' around $0.3-0.5$ in scale factor for ejecting
particles into large orbits.}
\label{fig:par_orbits_dist_EH_a}
\end{figure}

\subsection{Equation of Motion of an Ejected Particle}\label{sec:eq_of_motion}

The radial acceleration $\ddot{r}$ of an ejected particle, is to first
order, dominated by two terms: One from the gravitational field of the ejector halo and one
from the expanding background \citep{2012MNRAS.422.2931N, 2013JCAP...06..019B}. In this approximation,
the total acceleration is given by
\begin{equation}
\ddot{r}  = -\frac{GM_{1}}{r^{2}} -{H_0^{2} r}(\Omega_{m,0}a^{-3} - 2\Omega_{\Lambda, 0})/2,
\label{eq:par_acc_halo_exp}
\end{equation}
where $M_{1}$ is the time dependent mass of $H_{1}$, $r$ is the physical distance between
the center of $H_{1}$ and the particle, $H_{0}$ is the Hubble parameter
today, $a$ is the scale factor (not to be confused with the HQ scale radius), and $\Omega$ is the density parameter.
One can see that the force
exerted by the background expansion can either be attractive \emph{or} repulsive, depending on whether the universe is
decelerating or accelerating, respectively. The expansion itself does therefore not imply a repulsive force \citep{2003AmJPh..71..358D}.

The scale factor $a$ evolves in time by the standard relation \cite[e.g.][]{2013JCAP...06..019B}
\begin{equation}
a(t) = a_{m\Lambda}\left[\text{sinh}\left(3H_{0}t\sqrt{\Omega_{\Lambda,0}}/2\right)\right]^{2/3},
\label{eq:scalefac_t}
\end{equation}
where $a_{m\Lambda}$ is the matter - dark energy equality scale factor given by $(\Omega_{m,0}/\Omega_{\Lambda,0})^{1/3}$. 
The mass of the ejector halo $H_{1}$ is time dependent as well, due to matter accretion. To include this mass evolution,
we use the following empirical halo mass scaling \citep{2002ApJ...568...52W}
\begin{equation}
M_{1}(t) = M_{1}(a_{r})e^{\beta(z(a_{r}) - z(a))},
\label{eq:halomass_a}
\end{equation}
where $M_{1}(a_{r})$ is the mass of the halo at some reference scale factor $a_{r}$, $z$ is the redshift
and $\beta$ is a constant. The constant $\beta$ has been found to be in the range $0-2$, using numerical
simulations \citep{2002ApJ...568...52W, 2009MNRAS.398.1858M}.

\subsection{How Far Can an Ejected Particle Travel?}\label{sec:how_far_can_EP_travel}

The radial motion of a high velocity particle is found by solving
equation \eqref{eq:par_acc_halo_exp}, including equation \eqref{eq:scalefac_t}
for $a(t)$ and \eqref{eq:halomass_a} for $M(t)$.
We assume that the particle
escapes the virial radius of $H_{1}$ with a velocity $v_{eject}$ at a scale factor $a_{eject}$.
Results are shown in Figure \ref{fig:par_orbits_dist_EH_a}, which illustrates the particle's radial position $r(a)$
as a function of scale factor $a$, for different combinations of $v_{eject}$ and $a_{eject}$. The position is plotted in units
of the virial radius of $H_{1}$, which changes in time according to the 
defined relation $M_{\text{vir}} \equiv 4 \pi \Delta_{\text{vir}} \rho_{c} R_{\text{vir}}^{3} / 3 $,
where $\rho_{c}$ is the time dependent critical density of the universe and $\Delta_{\text{vir}}$ is the
overdensity threshold. The ejection velocity is given in units of
the corresponding virial velocity, defined
by $V_{\text{vir}} \equiv \sqrt{{GM_{\text{vir}}}/{R_{\text{vir}}}}$. We assume $\beta=1$, $\Delta_{\text{vir}}=200$,
and a flat universe with $\Omega_{m,0}=0.3$.
From the orbits shown in Figure \ref{fig:par_orbits_dist_EH_a}, we see that the maximum distance a particle can travel
strongly depends on its ejection time $a_{eject}$, as described in the following.

A particle ejected at early times will have a long time available to travel a long distance, but
will also experience a strongly increasing gravitational attraction from its ejector halo, due to mass accretion.
The force from the background is also attractive at early times, and the linear
dependence on the distance $r$ makes it therefore impossible for ejected particles to
become unbound \citep{2013JCAP...06..019B}. As a result of these effects, particles ejected early on
will quickly return to their ejector halo and are therefore unlikely to be found freely floating around today.

A particle ejected at later times will have less time to travel away from its ejector halo, but
will on the other hand experience a much smaller mass accretion, i.e. attracting force, from its ejector halo.
The force from the background also changes to be repulsive at late times,
which makes it even easier for late time ejected particles to escape.
As seen in Figure \ref{fig:par_orbits_dist_EH_a}, this interplay between ejection time and force terms results in a 'sweet spot'
around $a \sim 0.3-0.5$, for ejecting particles into large orbits.
The orbits also strongly depend on the ejection velocity $v_{eject}$. Dynamical kick mechanisms therefore play
a significant role in how matter distribute around cosmological halos.

An important observation from Figure \ref{fig:par_orbits_dist_EH_a},
is that particles ejected with only a few times the virial velocity
can enter large orbits, and travel several virial radii away from their target halo.
For example, a particle ejected at $a_{eject} \sim 0.5$ with
velocity $v_{eject} \sim 2V_{\text{vir}}$, will be $\sim 7R_{\text{vir}}$ away from the
ejector halo at present time.
Including dynamical effects for estimating how far particles can reach from their target halo is therefore important, and
leads to much higher limits compared to previous estimations based, e.g., on the halo collapse formalism \citep{2004A&A...414..445M}.
A similar conclusion was reached by, e.g., \cite{2007MNRAS.379.1475S, 2009ApJ...692..931L} using cosmological N-body simulations.

\section{Conclusions}\label{sec:conclusion}

We provide an explanation for how high energy particles are created in halo mergers,
by introducing a model which includes a double scattering mechanism.
The mechanism is a process where an incoming halo particle undergoes
two subsequent gravitational deflections during the merger, where the first is by the mass of the target
halo and the second is by the mass of the particle's original host halo.
The particle can receive a significant energy kick from this process,
because the two frames of deflection, i.e. the two halos,
move relative to each other during the merger.
The amount of energy generated through this mechanism, is comparable to
a well known energy contribution from tidal fields. The mechanism is therefore playing
a significant role in how energy is distributed in halo mergers. To our knowledge, the double scattering mechanism
has not been characterized in this context before, despite its great importance especially
for explaining the origin of high velocity particles.

From our presented model, we derive analytically the kick energy a given particle receives
from tidal fields and the double scattering mechanism,
as a function of its position in its original host halo just prior to the merger.
In the case of a $1:10$ head-on merger, we estimate that the largest energy kick is about $0.3-0.4\Phi_{1}(0)$, and is
given to particles located at around $\sim 0.1-0.2R_{\text{vir}}$ from the original host halo center.
We find this to be in agreement with numerical simulations.

By converting kick energy to velocity, we illustrate 
that our presented mechanisms can kick objects
to a resultant velocity $\sim 2$ times the virial
velocity of the target halo measured at its virial sphere.
This motivates us to suggest, that the high velocity of the recently discovered globular cluster
HVGC-1 \citep{2014ApJ...787L..11C}, can be explained by a halo merger, i.e., that
the energy kick is generated by tidal
fields and the double scattering mechanism.
We believe this serves as a more natural explanation compared to
other proposed ideas, including three-body interactions with a binary SMBH system in M87.
Cosmological simulations also support this \citep{2007MNRAS.379.1475S}.

Finally, from solving the equation of motion of a dynamically kicked particle in an expanding universe, we
find a 'sweet spot' around a scale factor of $0.3-0.5$ for ejecting particles into large orbits. These orbits can
easily reach beyond $\sim 5$ virial radii from the target halo, which is significantly longer than previous estimates based
on halo collapse models \cite[e.g.][]{2004A&A...414..445M}. This illustrates the importance of including dynamical
interactions for describing the outer regions of cosmological halos.

\acknowledgments{
It is pleasure to thank S. Hansen, T. Brinckmann, J. Zavala, R. Wojtak, M. Sparre, J. Hjorth, S. Pedersen, J. Fynbo, J. Zabl,
K. Finlator and S. Hoenig for comments and helpful discussions. 
The Dark Cosmology Centre is funded by the Danish National Research Foundation.
}

\bibliographystyle{apj}

\end{document}